\documentclass[a4paper,UKenglish]{lipics-v2021}

\nolinenumbers

\usepackage{todonotes}
\usepackage{tcolorbox}
\usepackage{booktabs}
\usepackage{array}
\usepackage{ragged2e}
\usepackage{tikz}
\newcolumntype{P}[1]{>{\RaggedRight\arraybackslash}p{#1}}

\title{Faster than the Team, Faster than the Customer:\newline Tool Integration, Collaboration, and Organisational Lag in AI-assisted RE}

\titlerunning{AI-assisted RE in Industrial Practice} 

\author{Jan-Philipp Stegh\"{o}fer}{XITASO GmbH Software \& IT Solutions, Augsburg, Germany}{jan-philipp.steghoefer@xitaso.com}{https://orcid.org/0000-0003-1694-0972}{}

\authorrunning{J.-P. Stegh\"{o}fer}

\Copyright{Jan-Philipp Stegh\"{o}fer}



\category{} 

\relatedversion{} 

\acknowledgements{I would like to thank the POs at XITASO for their support, particularly Lena Brehm for her support in planning this study and shaping its initial thrust.}

\EventEditors{TBD}
\EventNoEds{2}
\EventYear{2026}
\EventLogo{}
\SeriesVolume{TBD}
\ArticleNo{TBD}

\begin{document}

\maketitle

\begin{abstract}
The impact of applying generative AI tools to requirements engineering (RE) in industrial practice remains poorly understood.
This paper examines how AI-assisted RE tools are used in industrial practice at XITASO, a medium-sized enterprise for high-tech software engineering, and how they reshape workflows, tool integration, and PO--developer relationships.
We combine a 2024 company-wide use-case survey with two rounds of semi-structured interviews with eight product owners (POs) in late 2025 and spring 2026, covering an in-house chatbot and seven commercial AI tools.
We identify 15 distinct use cases across four categories: product backlog management, tender management, requirements and domain understanding, and document and artifact creation. Three findings emerge. First, the effect of AI on PO--developer interaction is mixed: the prevailing single-user interaction model can substitute for collaborative dialogue, and developers do not always welcome AI-generated artefacts. Second, tool integration\,--\,not tool capability\,--\,is the binding constraint: where integration is in place, time savings are dramatic; where it is missing, POs fall back on manual workarounds. Third, AI advances faster than the surrounding organisational systems, so its benefits accrue to individual POs while team processes and customer readiness remain the bottleneck.
The empirical GenAI-RE literature remains dominated by early-stage, lab-oriented evaluations of isolated tasks while practice has moved into territory it has not yet studied: practitioners are already assembling cross-tool integrations, navigating customer governance, and renegotiating role boundaries. From these patterns we derive a set of questions practitioners considering AI-assisted RE may ask of their own situation.
\end{abstract}

\section{Introduction}

Product owners and other practitioners are putting AI assistants to work on requirements engineering tasks every day. The tool landscape they face is heterogeneous and moving quickly: in-house GDPR-compliant chatbots, dedicated RE assistants integrated into platforms such as Jira (e.g., Atlassian's Rovo\footnote{\url{https://www.atlassian.com/software/jira/ai}}), and general-purpose AI tools wired into the existing toolchain via emerging protocols such as Model Context Protocol (MCP)\footnote{\url{https://modelcontextprotocol.io}}. What changes when these tools enter daily, multi-stakeholder RE practice is less clear.

\pagebreak
While AI-based coding assistants have received substantial empirical attention~\cite{fan2023large,weisz2025examining,chen2026beyond,wang2024trust}, work on AI-assisted requirements engineering (RE) in industrial practice is comparatively thin. Existing studies focus on isolated tasks in academic or controlled settings~\cite{cheng2026generative}, on specific techniques~\cite{guo2025natural}, or on GenAI adoption in software engineering broadly~\cite{russo2024navigating}. RE-specific industrial experience\,--\,what tools practitioners actually use, how they wire them into existing toolchains, and what changes in everyday RE work as a result\,--\,remains scarce.

In this paper, we examine how AI-assisted RE tools are used in practice at XITASO, a medium-sized enterprise (SME) for high-tech software engineering, and how they affect requirements engineering work over time. The longitudinal nature of the study lets us capture not only what these tools do but also how their use, their integration, and the workflows around them have changed in a comparatively short period.

We follow a qualitative approach combining a 2024 company-wide survey of relevant use cases with two rounds of semi-structured interviews with eight product owners (POs) at XITASO, conducted in late 2025 and spring 2026. Our study covers an in-house chatbot and seven commercial AI tools used for RE, only two of which are specifically designed as RE tools. We also contrast the resulting picture of the state of practice with the current GenAI-RE literature. This tackles our research question:

\begin{description}
    \item[RQ] How does AI-assisted requirements engineering impact requirements engineering work in a software development organisation?
\end{description}

The contribution of this paper is threefold: (1)~an overview of AI-assisted RE use cases and tools as deployed in industrial practice; (2)~empirically grounded findings on how AI-assisted RE tools impact requirements engineering work in a software development organisation, distilled into a set of practitioner-facing questions for self-assessment; and (3)~an empirical contrast between this practice and the current state of GenAI-RE research. This work provides industrial evidence for current research efforts on AI-assisted RE.

Three findings emerge. First, the effect of AI on the relationship between POs and developers is mixed: developers may not register the quality improvements POs make, may resist AI-generated artefacts emotionally, and the prevailing single-user dialogue model can substitute AI for the collaborative refinement that gives requirements their meaning. Second, tool integration, not tool capability, is what determines whether AI value is realised. Third, AI is reshaping workflows, role definitions, and patterns of communication faster than the surrounding organisations, including the organisation of the customer, can absorb the change. Together, these findings motivate the title of this paper: AI-assisted product owners are \emph{faster than the team, faster than the customer}.

The remainder of this paper is structured as follows. Section~\ref{sec:related-work} reviews related work; Section~\ref{sec:method} describes our research approach; Sections~\ref{sec:usecases} and~\ref{sec:tooling} present use cases and tools; Section~\ref{sec:discussion} presents findings and discusses implications; Section~\ref{sec:threats} describes the threats to validity and their mitigation; Section~\ref{sec:conclusion} closes the paper.

\section{Related Work}
\label{sec:related-work}

Several recent studies have evaluated LLMs for specific RE tasks. Ronanki et al.~\cite{ronanki2023investigating} investigated ChatGPT's potential for requirements elicitation, finding that it performs well on quality attributes compared to human RE experts. Santos et al.~\cite{santos2025adoption} surveyed practitioners about LLM adoption for user story generation, finding that two thirds already use LLMs for RE\,--\,primarily for brainstorming and content review, but report concerns about lack of context and over-reliance. Hassani et al.~\cite{hassani2025empirical} evaluated LLMs for classifying requirements-relevant provisions in regulatory documents, achieving high precision with GPT-4o. However, these studies evaluate isolated tasks in controlled or survey-based settings; our work extends this by studying how AI-assisted RE tools function in daily industrial practice across multiple use cases and within real development workflows.

At the organisational level, several studies examine how software teams adopt GenAI tools. Kemell et al.~\cite{kemell2025still} conduct a multiple case study asking whether GenAI tools have moved beyond personal assistant roles in software organisations. Simaremare and Edison~\cite{simaremare2024state} survey experienced practitioners on adoption patterns, benefits, and challenges. Russo~\cite{russo2024navigating} examines the complexity of GenAI adoption broadly. Our work narrows the focus to RE specifically and provides longitudinal evidence showing that tools are indeed moving past the personal assistant stage through integrations like MCP, but that this creates new challenges for team collaboration.

For document-intensive RE tasks, Nai et al.~\cite{nai2024large} demonstrate LLMs' potential for analysing public procurement documents, the same problem domain addressed in our study. On the human-AI interaction side, two strands of work frame how we read our data. Seeber et al.~\cite{seeber2020machines} sketch a research agenda for AI as a teammate, arguing that integrating AI into team workflows changes the nature of the team itself. Webber~\cite{webber2024paradox} examines the paradox of AI as teammate, arguing that AI can simultaneously improve individual efficiency while undermining team dynamics. Storey~\cite{storey2026intentdebt} more recently introduces a triple debt model. It contains \emph{cognitive debt}, the team-level erosion of shared understanding as AI-generated artefacts accumulate faster than the team can build a common mental model, often manifesting as a loss of ``transactive memory'' in which team members lose track of who knows what and increasingly rely on individual interactions with AI rather than shared practices. It also contains \emph{intent debt}, the absence or erosion of explicit rationale, goals, and constraints that humans and AI agents need to evolve software safely. Our findings provide empirical evidence for cognitive debt in the RE domain: improved tool integration reduces the need for developer involvement, potentially deepening the collaboration gap rather than bridging it as the team loses the conversational context behind AI-generated artefacts, and the workflows and roles around AI use are absorbing change more slowly than AI itself is being adopted.

\section{Methodology}
\label{sec:method}

Following the criteria of Wohlin and Rainer~\cite{wohlin2022case}, our study is best characterised as a qualitative interview study with a survey as a starting point, rather than a case study, since it lacks strong data triangulation of data sources. We follow the empirical standards for software engineering research~\cite{ralph2020empirical}, in particular the guidelines for interview studies.

\paragraph*{Case Description}
XITASO is a medium-sized software engineering service provider founded in 2011. In early 2026, it has 240 employees across nine offices in Germany and three in Spain with its headquarter in Augsburg, Germany. It offers high-end software engineering services to customers mainly in the machinery and production engineering (50\% of revenue), medical devices (30\% of revenue), and public sector (10\% of revenue), with other customers in insurance, energy, and elsewhere. It is a network organisation, meaning that there are no strict reporting hierarchies and the nucleus of most decisions, including who can go on vacation when, is the team. Cross-cutting concerns are captured in communities of practice, of which there are over 30, including technical communities for topics such as AI, software engineering, and user experience, as well as communities focused on shared services, such as the office environment, personnel development, or finances. 

Development teams consist of 5 to 15 people, with the median around 7 people. These teams usually consist of one PO, one Scrum Master, and several developers, with support from AI engineers, security engineers, or user experience experts where warranted. One team usually works on one customer project at a time; this rule is sometimes broken for short projects that last less than 8 weeks. Rather than creating new teams for such short projects, a second or sometimes a third project is tackled by an established large team with free capacity. Teams are usually formed based on domain experience of the personnel or specific technical abilities or certifications. For instance, there are teams that have specific experience in robotics which are kept together. 

All development teams work in close collaboration with customers on requirements engineering as part of agile development projects. Most projects are ``time and material'' with a certain budget which means that the outcome of the project is shaped in an ongoing discussion with the customer. 

\paragraph*{Data Collection}

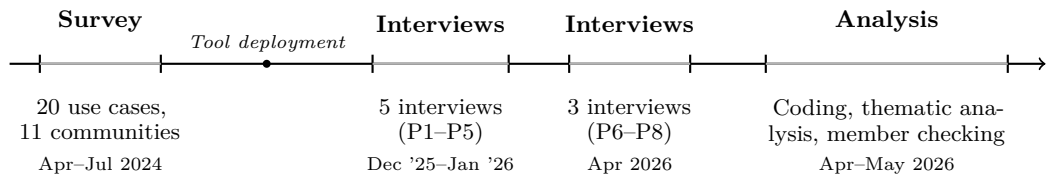
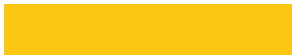
\begin{figure}[t]
\centering
\begin{tikzpicture}[
    phaselabel/.style={font=\small\bfseries, above=0.3cm, align=center},
    datalabel/.style={font=\footnotesize, below=0.35cm, align=center, text width=2.4cm},
    timelabel/.style={font=\scriptsize}
]
\draw[->, thick] (-0.2,0) -- (13.5,0);

\draw[thick] (0.2,-0.12) -- (0.2,0.12);
\draw[thick] (1.8,-0.12) -- (1.8,0.12);
\draw[thick, gray!60] (0.2,0) -- (1.8,0);
\node[phaselabel] at (1,0) {Survey};
\node[datalabel] at (1,0) {20 use cases, 11~communities};
\node[timelabel, below=1.1cm] at (1,0) {Apr--Jul 2024};

\node[font=\scriptsize\itshape, above=0.04cm] at (3.2,0) {Tool deployment};
\fill (3.2,0) circle (1.5pt);

\draw[thick] (4.6,-0.12) -- (4.6,0.12);
\draw[thick] (6.4,-0.12) -- (6.4,0.12);
\draw[thick, gray!60] (4.6,0) -- (6.4,0);
\node[phaselabel] at (5.5,0) {Interviews};
\node[datalabel] at (5.5,0) {5 interviews (P1--P5)};
\node[timelabel, below=1.1cm] at (5.5,0) {Dec '25--Jan '26};

\draw[thick] (7.2,-0.12) -- (7.2,0.12);
\draw[thick] (8.8,-0.12) -- (8.8,0.12);
\draw[thick, gray!60] (7.2,0) -- (8.8,0);
\node[phaselabel] at (8.0,0) {Interviews};
\node[datalabel] at (8.0,0) {3 interviews (P6--P8)};
\node[timelabel, below=1.1cm] at (8.0,0) {Apr 2026};

\draw[thick] (9.8,-0.12) -- (9.8,0.12);
\draw[thick] (13.0,-0.12) -- (13.0,0.12);
\draw[thick, gray!60] (9.8,0) -- (13.0,0);
\node[phaselabel] at (11.4,0) {Analysis};
\node[datalabel, text width=3.4cm] at (11.4,0) {Coding, thematic analysis, member checking};
\node[timelabel, below=1.1cm] at (11.4,0) {Apr--May 2026};

\end{tikzpicture}
\caption{Overview of the research approach and timeline.}
\label{fig:methodology}
\end{figure}

Our study followed a three-phase data-collection design: an initial survey, followed by two rounds of semi-structured interviews at different points in time. Analysis was conducted iteratively across all three phases (see Figure~\ref{fig:methodology} for an overview).

Between April and July of 2024, we conducted a \emph{company-wide survey} across XITASO's communities of practice to identify relevant use cases for AI assistance. Each community was asked to fill in a use case template on a wiki page. The template contained three questions to answer: ``What’s your pain point?'', ``How can GenAI address it?'', and ``Which tools do you know that could help?''. The communities of practice discussed the use cases in their internal meetings and updated the wiki page directly. In total, we collected 20 use cases from 11 communities. A committee of four subject matter experts from the company then selected the two use-case categories with the strongest relation to requirements engineering activities and the highest potential time savings and task complexity as criteria; these two categories (Product Backlog Management and Tender Management) were subsequently refined into 15 more fine-grained use cases through the interviews (see Section~\ref{sec:usecases}). To address these initial categories, we identified several commercially available and in-house AI-assisted RE tools and evaluated their adoption across the company (see Section~\ref{sec:tooling}).

We subsequently conducted \emph{semi-structured interviews with eight product owners} at XITASO who had hands-on experience with the deployed AI-assisted RE tools in their daily work. This constitutes purposive sampling: we interviewed those eight out of the 17 product owners who were actively using AI based on recommendations by their peers and who have significant experience as POs of at least five years. Their characteristics are shown in Table~\ref{tab:interviewees}. The experience requirement serves as a control to ensure that POs have a foundation to understand the impact of AI tools on their daily work and on collaborative workflows. POs that do not use AI do not necessarily do that out of personal preference: some of our customers expressly forbid the use of AI or limit the available tools severely.

The interviews were conducted in two phases: the first five were conducted in December 2025 and January 2026, three more in April 2026. This allows us to also introduce a longitudinal perspective and track how the use of AI tools has changed during that period of time. Both sets of interviews were conducted using the same interview guide which is available in the supplementary material~\cite{supplementary-material}. All interviews were conducted online and recorded. The language was German. The interviewer took extensive notes in German that were validated and completed based on the recordings.

\vspace{1em}The interviews were conducted by the author, who has over ten years of experience in empirical software engineering research, with a focus on requirements engineering, agile software development, and model-driven engineering. Originally holding an academic position, the author now works in industry as Head of Research and Innovation at the organisation in which this study was conducted.
The author is therefore a partial insider: the participants are colleagues within the same organisation, though from different teams and without a day-to-day working relationship. This insider position offered advantages, including domain familiarity, shared vocabulary, and established trust, but also introduced potential risks. In particular, the author's senior organisational role may have influenced participants' willingness to share critical or dissenting views, despite assurances of confidentiality.

To mitigate these risks, the following measures were taken: Confidentiality was explicitly assured to all participants prior to the interview, and all findings reported in this paper are anonymised so that no response can be attributed to an individual. Member checking was conducted, allowing participants to review and validate the interpretations drawn from their responses. An LLM-assisted review of the coded findings was performed as an additional analytical check, to surface potential blind spots or inconsistencies in interpretation that may have stemmed from the author's familiarity with the context.

\begin{table}[t]
\caption{Characteristics of interviewed product owners.}
\label{tab:interviewees}
\centering
\begin{tabular}{@{}llP{3.4cm}cP{4.5cm}l@{}}
\toprule
\textbf{ID} & \textbf{Exp.} & \textbf{Domain} & \textbf{Team Sz.} & \textbf{Tool(s) Used} & \textbf{Usage} \\
\midrule
P1 & 5 yrs. & Machinery & $\sim10$ & ChatXiPT, Product Copilot, Claude Desktop & 8 mts.\\ 
P2 & 7 yrs. & Medical Technology, Healthcare & $\sim15$ & ChatXiPT, Product Copilot & >2 yrs.\\ 
P3 & 10 yrs. & Machinery & 5 & ChatXiPT, Product Copilot & >2~yrs. \\ 
P4 & 5 yrs. & Public Sector, Healthcare & 3--8 & ChatXiPT, Product Copilot, TenderZen & >2~yrs. \\ 
P5 & 5 yrs. & Public Sector & $\sim15$ & ChatXiPT, Product Copilot, Claude Desktop & 1 yr. \\ 
\midrule
P6 & 10 yrs. & Insurance & $\sim13$ & MS Copilot, ChatXiPT & 2 yrs. \\ 
P7 & 18 yrs. & Ecommerce, Machinery & 5--8 & ChatXiPT, Claude Code & 1 yr \\ 
P8 & 8 yrs. & Public Sector & 7 & ChatXiPT, MS Copilot, TenderZen & >2 yrs.\\ 
\bottomrule
\end{tabular}
\end{table}

\paragraph*{Data Analysis}
We extracted individual data points about tool usage, perceived limitations, and opportunities from the interview notes. Where necessary, data points were translated into English using DeepL Pro and collected on an online whiteboard. We combined initial structural coding~\cite{namey2008data} based on the interview guide with thematic analysis~\cite{clarke2017thematic}. Each data point was recorded as a card; cards addressing the same topic but distinct points were connected, with codes added as labels. Sticky notes denoted specific use cases. We used Claude Sonnet 4.6 to derive coding guidelines and identify inconsistencies, which were manually reviewed and resolved over several iterations. We employed member checking~\cite{lincoln1985naturalistic} to validate the data: interview summaries were sent to participants, who later also had the opportunity to review the main findings and provide feedback.


\vspace{1em}The creation of this paper was supported by AI. Claude Sonnet 4.6 and Claude Opus 4.6 were used to critique drafts of the manuscript, brainstorm ways to present this work and frame the research questions, summarize literature, and for editorial work. In addition, Claude Opus 4.6 was used to validate the coding and the derived findings based on the notes and it created some of the text passages in this work. All AI-assisted outputs were critically reviewed, validated, and revised by the author. The author retains full responsibility for the research design, analytical decisions, interpretations, and conclusions.

\section{Use Cases for AI-assisted Requirements Engineering}
\label{sec:usecases}

Based on the survey, we initially identified two main areas for AI-assisted RE at XITASO: \emph{Product Backlog Management} and \emph{Tender Management}. The interviews revealed each to be a whole category of fine-grained use cases, and uncovered two further categories: \emph{Requirements \& Domain Understanding} and \emph{Document \& Artifact Creation}. In total, we identified 15 use cases across these four categories. Table~\ref{tab:mapping} provides the complete mapping with participants. Below, we focus on the most widely practised and analytically consequential use cases.

\paragraph*{Product Backlog Management}
Product backlog management is a major task of our POs. Writing high-quality user stories in rapidly changing projects is difficult, as is editing user stories or bug reports authored by other team members or the customer. Backlog maintenance and cleanup consume considerable time, while identifying and linking relevant items remains a persistent challenge. A clean backlog with well-defined backlog items is, however, essential for good planning and to enable developers to work productively~\cite{lucassen2016improving}.

\emph{Backlog Refinement} is the most widely practised use case, mentioned by seven of eight POs. GenAI tools support a range of cleanup and enrichment tasks: splitting epics, finding duplicates, adding or improving acceptance criteria, and identifying missing edge cases or gaps in individual PBIs. Notably, prioritisation and effort estimation are exceptions: current tools handle neither reliably, and POs continue to do these manually. Tools draw on project-specific knowledge to propose standardised acceptance criteria phrases, flag stories missing critical information, and detect dependencies. At XITASO, Product Copilot, ChatXiPT, and MS Copilot are all used for this purpose — the former two for Jira-based projects (P1, P2, P3, P5, P8), and MS Copilot for an Azure DevOps-based project (P6).

A newer practice, which we categorize as \emph{implementation-aware} Backlog Refinement (P7), goes beyond standard backlog refinement. P7 uses Claude Code, connected via MCP server to the project's Jira board and source code repository (see Section~\ref{sec:tooling}), to elaborate tickets in the full context of the codebase. The PO converses with the AI to: (1)~define technical constraints informed by the actual architecture, (2)~assess the feasibility of requirements against the current implementation, and (3)~draft acceptance criteria reflecting real system behaviour. Rather than merely structuring requirements, the AI provides technical context previously accessible only through developer conversations. This allows the PO to ask ``is this requirement feasible given our current architecture?'' without waiting for a refinement session. This blurs the boundary between requirements engineering and implementation, with direct consequences for collaboration patterns (see Section~\ref{sec:discussion}). P3 expressed a closely related desire in late 2025: ``I would like to connect my LLM to a `staging area' to better understand a requirement.''

\emph{Story Authoring} (P2, P3, P5, P6) and \emph{Backlog Bootstrapping} (P1, P2, P3, P8) both address the creation of backlog items. In Story Authoring, POs use Product Copilot or ChatXiPT to generate individual PBIs from notes or prose, filling in templates, suggesting acceptance criteria, and surfacing CRUD and validation scenarios. Backlog Bootstrapping goes further: given bulk input such as an Excel spreadsheet, screenshots, or a specification document, Product Copilot generates a large set of PBIs in one step — with multiple participants noting dramatic time savings over manual creation. \emph{Backlog Search} (P2; aspirational for P6) is currently a niche case, enabling semantically relevant stories to be retrieved for a given query, e.g., to assess the impact of an architectural decision across the backlog.

\paragraph*{Tender Management}
Screening tender documents and extracting relevant information requires significant resources. POs must make rapid bid/no-bid decisions, extract requirements to estimate effort and price, and identify suitable reference projects and personnel profiles. At XITASO, tender management follows a four-stage pipeline, used primarily by P4 and P8.

\emph{Tender Discovery}: TenderZen agents search tender portals, upload documents, and pre-filter results based on fit with company capabilities and desired areas of work, reducing the manual effort of identifying relevant opportunities (P4, P8).

\emph{Tender Analysis} is the central stage of the pipeline. TenderZen ingests a tender document and extracts requirements, deliverables, and quality criteria into a structured form, while identifying gaps between tender requirements and company capabilities. Rather than manually searching the documents, POs can query the content interactively via chat or pre-defined templates to efficiently retrieve critical information such as tender volume, submission deadlines, or required documentation. ChatXiPT and MS Copilot are also used for targeted document queries. A known limitation flagged by P8 is that TenderZen's analyses do not reference original data sources, complicating verification of extracted content.

\emph{Reference Matching} (P8) uses TenderZen and Rovo to find past projects and personnel profiles in Confluence that match the requirements of a given tender, and to adapt reference descriptions to the tender context. The handoff between Rovo and TenderZen is currently manual (see Section~\ref{sec:discussion}). 

\emph{Bid Authoring} (P4, P8) closes the pipeline. ChatXiPT and MS Copilot are used to draft and optimise bid concepts and proposals; POs find AI more effective for reviewing and optimising existing content than for creating bids from scratch. Product Copilot fulfils a more specific role: it translates the requirements extracted during Tender Analysis into a product backlog (an instance of the \emph{Backlog Bootstrapping} use case described above), which is then refined and estimated using the standard backlog-management workflow.

\paragraph*{Document and Artifact Creation}
\emph{UI Prototyping} (P1, P3, P5) is the primary use case of this category. Rapid prototyping of parts of user interfaces supports understanding the solution space for a requirement as well as gathering early feedback from customers. Simple, clickable UIs make it possible to test whether a workflow is feasible, whether all data is available, and how interface elements need to be laid out\,--\,providing a concrete artifact to discuss rather than an abstract description spread across several user stories. Chatbots expose this capability through \emph{artifacts}: interactive HTML prototypes generated from prompts and viewable directly in the chat window. Providing an example of the current design language\,--\,e.g., as a screenshot\,--\,allows the AI to style the prototype accordingly, and modifications to the artifact are reflected immediately in the requirements. POs can prototype live during customer discussions or backlog grooming sessions. Our POs initially used Claude Desktop for this purpose; since February 2026 this feature is also available in ChatXiPT.

The remaining use cases in this category are practised by fewer participants. \emph{Document Creation \& Optimisation} (P1, P3, P8) covers drafting, converting, and polishing documents and presentations, as well as extracting requirements into structured formats such as Excel. P1 and P3 use ChatXiPT for this purpose; P8 also uses MS Copilot for tasks tied to the Microsoft 365 environment. \emph{Documentation Generation} (P5) derives user-friendly product descriptions from requirements and code; P5 uses Claude Desktop, having tested Scribe and found it inferior. \emph{Change Summarisation} (P5, P6, P7) is currently a desired rather than an established practice: P5 tested Product Copilot for summarising product changes and ticket discussions in support of prioritisation and onboarding, but found it not viable.

\paragraph*{Requirements and Domain Understanding}
\emph{Vision \& Roadmap Derivation} (P2, P7) addresses a challenge that arises early in product development: arriving at a shared, articulate product vision before detailed requirements exist~\cite{ebert2006understanding}. P2 uses ChatXiPT and Product Copilot to derive a product vision, target audience, and added value from existing artefacts such as a Jira board, wiki pages, or a requirements document. The resulting vision statement serves as a starting point for solution concepts and customer discussions. P7 uses a similar approach to define how the product vision should evolve over time. This use case is particularly valuable in early project phases where the direction is still being negotiated.

\emph{Domain Exploration} (P3, P6, P7) uses ChatXiPT and MS Copilot as a faster alternative to traditional search for understanding domain terminology, evaluating technology options, and brainstorming solution approaches\,--\,effectively accelerating the contextualisation that precedes requirements work. \emph{Interview \& Meeting Analysis} is used by P7 via Curly, an AI-powered interview bot that conducts structured stakeholder interviews, transcribes responses, and summarises findings by question; P3 expressed this as a strong future desire. POs note a split in stakeholder acceptance: technically experienced interviewees embrace AI-mediated interviews, while others are hesitant.






\begin{table}[p]
\caption{Mapping of use cases to AI-assisted RE tools at XITASO. The * indicates a desired use case, $^\dagger$ that a tool was experimented with but not adopted.}
\label{tab:mapping}
\centering
\begin{tabular}{@{}P{2.6cm}P{3cm}P{5.4cm}P{1.6cm}@{}}
\toprule
\textbf{Use Case} & \textbf{Tool(s)} & \textbf{Key Capability} & \textbf{Part.} \\
\midrule
\multicolumn{4}{@{}l}{\textbf{Product Backlog Management}}\\
\midrule
    Story Authoring &
    Product Copilot, ChatXiPT, MS Copilot &
    Fill in story templates from notes or prose; suggest acceptance criteria, validation, and CRUD operations &
    P2, P3, P5, P6 \\

    Backlog Bootstrapping &
    Product Copilot &
    Generate multiple PBIs from Excel or specification documents in bulk &
    P1, P2, P3, P8 \\

    Backlog Refinement &
    Product Copilot, MS Copilot, ChatXiPT, Claude Code &
    Sort, prioritize, split backlog items; find duplicates; add acceptance criteria; identify gaps &
    P1, P2, P3, P5, P6, P7, P8 \\

    Backlog Search &
    Product Copilot &
    Semantically query a backlog to find stories relevant to a given topic or architectural decision &
    P2, P6* \\
\midrule
\multicolumn{4}{@{}l}{\textbf{Tender Management}}\\
\midrule

    Tender Discovery &
    TenderZen &
    Search tender portals, upload documents, and pre-filter based on fit &
    P4, P8 \\

    Tender Analysis &
    TenderZen, ChatXiPT, MS Copilot &
    Extract requirements, deliverables, and quality criteria from tender documents; identify capability gaps &
    P4, P8 \\

    Reference Matching &
    TenderZen, Rovo &
    Match tender requirements to existing reference projects and personnel profiles stored in Confluence &
    P8 \\

    Bid Authoring &
    ChatXiPT, Product Copilot, MS Copilot &
    Develop and optimize bid concepts and proposals; review against previous bid feedback &
    P4, P8 \\
\midrule
\multicolumn{4}{@{}l}{\textbf{Document and Artifact Creation}}\\
\midrule

    Document Creation \& Optimization &
    ChatXiPT, MS Copilot &
    Draft, convert, optimize documents; extract requirements into structured formats such as Excel &
    P1, P3, P8 \\

    UI Prototyping &
    Claude Desktop &
    Generate interactive UI prototypes from screenshots or specifications &
    P1, P3, P5 \\

    Documentation Generation &
    Claude Desktop &
    Derive user-friendly product descriptions from requirements and code &
    P5 \\

    Change Summarization &
    Product Copilot$^\dagger$ &
    Summarize product changes and ticket discussions to support prioritization and onboarding &
    P5*, P6*, P7* \\
\midrule
\multicolumn{4}{@{}l}{\textbf{Requirements and Domain Understanding}}\\
\midrule
    Domain Exploration &
    ChatXiPT, MS Copilot &
    Research domain terminology, evaluate technology options, and brainstorm solution approaches &
    P3, P6, P7 \\

    Interview \& Meeting Analysis &
    Curly &
    Conduct structured stakeholder interviews; transcribe, summarize, and build knowledge graphs &
    P3*, P7 \\

    Vision \& Roadmap Derivation &
    ChatXiPT, Product Copilot &
    Extract product vision, target audience, and added value from artifacts &
    P2, P7 \\
    \bottomrule
\end{tabular}
\end{table}

\section{Tooling for AI-assisted Requirements Engineering at XITASO}
\label{sec:tooling}

Several distinct AI tools surfaced as part of RE practice across the eight POs we interviewed. They divide into two broad groups: \emph{dedicated RE tools} built specifically for backlog or tender management (Product Copilot, TenderZen), and \emph{general-purpose AI tools} applied to RE tasks (ChatXiPT, Claude Code, Claude Desktop, MS Copilot, Curly, Rovo). ChatXiPT is an in-house development built on an open-source platform; all other tools are commercially available. Table~\ref{tab:mapping} provides the full mapping of tools to use cases. All descriptions are a snapshot at the time of writing.

\paragraph*{ChatXiPT} 
We use our own GDPR-conformant chatbot based on LibreChat\footnote{\url{https://www.librechat.ai/}} which provides access to different frontier models, including GPT-5 and Claude Sonnet. It has been first deployed and released to the company in September 2023. All eight POs we interviewed use this chatbot for tasks that require little context: refining user stories, brainstorming new features, and validating the technical viability of an idea. These tasks can be done in the conversational interface after copying and pasting a limited amount of data. The LLM partially replaces a human colleague and allows the PO to get feedback\,--\,in particular with prompting to ensure that the LLM is critical and provides constructive feedback. Since February 2026, ChatXiPT also creates user interface prototypes for a set of requirements to collect rapid feedback from customers (see Section~\ref{sec:usecases}), which made Claude Desktop for the same purpose obsolete.

We extended LibreChat with a custom RAG pipeline to analyse complex documents and integrate their content into the responses of the LLM. This feature is used by our POs to analyse complex technical documentation and derive requirements from them. One example of such technical documentation are the documents that are part of public tenders, but this work has been replaced by TenderZen (see below) which offers a more structured interface for that use case.

\paragraph*{Product Copilot}
Product Copilot\footnote{\url{https://product-copilot.ai/}} is an AI assistant integrated into Jira that supports POs in creating and refining backlog items. The tool draws on the existing backlog, wiki pages, a product description, and user-maintained history as context, ensuring that generated backlog items are based on existing project information. It is possible to tailor the writing style and use existing templates to fit a project's conventions. A browser plugin integration avoids manual copying of text between tools. Product Copilot has been used in different projects at XITASO since July 2025.

Five POs (P1--P5) appreciate the tool for filling in details like personas and acceptance criteria, but find it most valuable for less experienced POs. Product Copilot can derive requirements from screenshots or text files, and structure large requirements into epics. It identifies ambiguous backlog items and missing elements, which are less critical in mature teams. Some POs prefer chatting in a colloquial tone and having the tool formalise their input into a well-structured backlog item, while experienced POs often feel they can write better stories directly. The tool also extracts condensed artefacts such as a product vision or the main target audiences from the backlog (see \emph{Vision \& Roadmap Derivation} in Section~\ref{sec:usecases}), which helps POs communicate with customers and validate the backlog.

\paragraph*{TenderZen}

TenderZen\footnote{\url{https://www.tenderzen.de/en}} is an AI-powered tender management platform that searches procurement platforms across Europe and Germany, including local and municipal sources. The tool supports keyword and profile-based searches and provides analysis features for tender documents uploaded as ZIP files, generating reports that extract core requirements, technical and quality criteria, and candidate reference projects from company profiles. Analyses are shared across users of the platform. Custom chat sessions allow POs to query additional information from the extracted content. The platform supports matching manually uploaded employee profiles to tender requirements. TenderZen has been in use at XITASO since September 2025.

At XITASO, the tender team (P4 and P8) processes approximately 30 tenders daily, spending about one hour on pre-selection and 1.5--2 hours examining 3--5 tenders in detail; a complex tender can consist of up to 30 documents of 20--30 pages each. TenderZen has reduced the evaluation time for a single tender from four hours to 30 minutes. References must currently be uploaded manually, with API integration planned. P8 notes that TenderZen does not reference the underlying source material in its analyses, which complicates verification of extracted content.

\paragraph*{Claude Code}

Among the general-purpose tools, one PO (P7) has adopted Claude Code as a deeply integrated RE assistant, as described in Section~\ref{sec:usecases}. Claude Code uses MCP (Model Context Protocol)\footnote{MCP is an open protocol that enables AI assistants to connect to external tools and data sources such as issue trackers, code repositories, and documentation systems.} to connect to the project's Jira board and source code repository simultaneously. Unlike Product Copilot's browser plugin or ChatXiPT's copy-paste workflow, this integration gives the AI direct access to both the requirements artefact and the codebase. This pattern emerged in the second round of interviews (April 2026) and was absent in the first round (December 2025/January 2026), illustrating how rapidly both tool capabilities (vendor MCP support) and practitioner skills (configuring these integrations) are evolving. Claude Code has been in use at XITASO since February 2026.

\paragraph*{Claude Desktop}
Claude Desktop (Anthropic) was used by P1 and P5 as a standalone AI assistant for exploratory and creative RE tasks. Unlike Claude Code, it has no integration with the project toolchain. Its primary application has been UI prototyping and brainstorming. Since ChatXiPT gained UI artifact generation in February 2026, Claude Desktop's role at XITASO has diminished.

\paragraph*{MS Copilot}
Microsoft Copilot is used in two distinct contexts at XITASO. P8 uses it for office tasks: drafting and optimising proposals, presentations, and other documents in the Microsoft 365 environment, as well as converting requirements into structured Excel formats. P6 uses MS Copilot in the Microsoft 365 environment as well, alongside ChatXiPT, for tasks such as document drafting, requirement extraction, and domain onboarding. However, the customer P6 works for has not enabled MS Copilot's integration with Azure DevOps, which is the backlog system in this project. Because Product Copilot is Jira-focused and likewise does not apply to ADO, P6 has no AI tool that operates directly on the backlog\,--\,a missing integration the PO repeatedly identified as a source of friction.

\paragraph*{Curly}
Curly\footnote{\url{https://www.curly.so/}} is a voice- and text-based AI interview bot used by P7 for structured stakeholder elicitation at scale. The PO defines a questionnaire and sets the probing depth; Curly then conducts the interview, transcribes responses, and summarises findings by question, enabling requirements to be gathered from many stakeholders simultaneously (see \emph{Interview \& Meeting Analysis} in Section~\ref{sec:usecases}). Acceptance is split: technically experienced interviewees engage readily with AI-mediated interviews, while others are hesitant, limiting the current reach of the approach.

\paragraph*{Rovo}
Atlassian Rovo is an AI-powered knowledge search and content management tool integrated into Confluence. P8 uses it to find and adapt reference project descriptions stored in Confluence in support of the \emph{Reference Matching} use case (see Section~\ref{sec:usecases}): Rovo retrieves past project descriptions that match the requirements of a given tender, which are then adapted for the bid. Its scope is limited to content in the Atlassian ecosystem.

\paragraph*{Other tools}
Two further tools appeared in peripheral roles. \emph{Scribe} was tested by P5 for creating manuals and visualising user flows but was found inferior to Claude Desktop for this purpose and is not in active use. \emph{Figma Make} was used by a developer at P7's direction to generate UI wireframes and interactive prototypes from requirements; P5 also mentioned it as a potential alternative to Claude Desktop for sharing prototypes collaboratively, though it was not adopted directly by POs.

\section{Findings and Discussion}
\label{sec:discussion}

As AI tools increase developer productivity, the pressure on requirements engineers to fill backlogs faster grows accordingly. Yet RE fundamentally involves finding consensus through negotiation and conversation\,--\,a deeply human activity with natural limits on acceleration. The interviews with our eight POs reveal that AI is reshaping this activity along three interconnected dimensions: how AI affects the relationship between POs and developers; how tool integration (and its absence) determines whether AI assistance delivers real value; and how AI capabilities are advancing faster than the human and organisational systems needed to absorb them. This gap between AI capability and organisational capacity frames the findings below.

\paragraph*{AI changes the quality and character of developer--PO interaction}

The most immediate effect of AI-assisted RE is on the artefacts used by both POs and developers. AI tools reliably raise the floor of story quality\,--\,better formatted acceptance criteria, fewer omissions, more consistent structure\,--\,against the dimensions in which RE quality is conventionally evaluated~\cite{frattini2023requirements}. However, the data from the interviews reveal that this quality improvement does not translate automatically into better developer reception.

A key finding is that AI-generated stories are not always well received. P1 described a situation where a PO uploaded an AI-generated story without reviewing it: \emph{``In one case, I had already uploaded it without checking, and the developer reacted negatively to the highly technical, generated description.''} P5 observed resistance where one developer reacted with negativity to all tickets or even conversations with AI tools\,--\,\emph{``purely based on emotion, no rational arguments.''} This emotional resistance is a real phenomenon that AI adoption strategies must account for~\cite{wang2024trust}.

A second dimension is perception asymmetry. P1 noted that most developers do not necessarily notice the difference even though Product Copilot generates numbered acceptance criteria, while the PO sees it as a critical point. P3 confirmed this from the opposite side: developers have not provided any specific feedback on AI-generated stories; the PO reviewed all of them and revised as needed. POs may therefore be over-investing effort in AI-assisted quality improvements that developers barely register.

There is a deeper risk beyond reception: AI can substitute for the dialogue that shapes requirements. P5 identified this directly in the context of backlog refinement: \emph{``writing a story is based on a conversation with [Product Copilot] in which a lot of context is provided''}. The result is a story that needs to be discussed with the team\,--\,\emph{``in the future, this discussion could also happen between the dev team and the Copilot.''} P7 goes even further, using Claude Code to refine the technical aspects of a backlog item against the actual source code, which means that developers are no longer involved in shaping the understanding of the requirement and possible solutions. The literature on AI as a teammate has long observed that integrating AI into team workflows changes the team itself~\cite{seeber2020machines,webber2024paradox}, with consequences for shared understanding and accountability. The single-user interaction model of current tools amplifies this risk, as detailed below.

The current single-user model of AI-assisted RE tools amplifies this risk. All tools we observed\,--\,including ChatXiPT, Product Copilot, and TenderZen\,--\,follow a fundamentally individual interaction model: one user engages in a dialogue with the AI, and the team receives the resulting artefact but not the conversational context that shaped it. The two contexts\,--\,AI-generated and human-discussed\,--\,can drift apart, forcing the PO to manually bridge them by restarting the AI dialogue with discussion results or reconstructing AI reasoning for the team. Backlog refinement always requires discussion and negotiation regardless of how stories were produced; the AI artefact is the starting point, not the end point, of that conversation. Beyond information exchange, refinement sessions build shared understanding, surface assumptions, and create collective ownership of the work~\cite{verwijs2023theory}\,--\,none of which the AI currently participates in. The reliability of AI-generated artefacts adds another dimension: LLM-based tools can produce plausible but incorrect requirements (e.g., P2 identified a generated story that incorrectly assumed a user must register on every login), requiring additional review effort that partially offsets productivity gains~\cite{fan2023large,liu2024exploring}. 

Storey's framework~\cite{storey2026intentdebt} identifies two distinct debt types that accumulate in our setting. Between PO and AI, \emph{intent debt} accrues whenever a generated story is committed to the backlog without close review\,--\,P1's example of an unreviewed, overly technical story reaching a developer illustrates this: the artefact lacks the captured rationale and constraints that would allow the team to understand why it was written the way it was. Between the PO and the team, \emph{cognitive debt} accrues because the conversational context that shaped the story is invisible to developers in refinement\,--\,a loss of ``transactive memory'' in which the team inherits the artefact but not the shared understanding that produced it. The same is true for reliance on the PO's private AI dialogue rather than on the collaborative practices that spread that understanding, as in the case of P7. RE research acknowledges the social and negotiation aspects of requirements work~\cite{goguen1994requirements,damian2003challenges}; current GenAI-RE evaluations, by contrast, frequently reduce RE to a single-engineer scenario~\cite{ellsel2025advancing,krishna2024using}, missing precisely these dynamics.

\paragraph*{Tool integration is a major bottleneck}

Across nearly all participants, the most persistent frustration is not with AI capability but with the lack of integration between AI tools and existing workflows. When AI tools cannot access project context, their outputs are generic, irrelevant, or require significant manual effort\,--\,negating the productivity gains.

P3 described a fundamental structural problem: \emph{``the project is created separately in Product Copilot; it should be integrated directly into our standard project management workflow.''} P6 was more direct: \emph{``without a connection between Copilot and ADO, context is missing''} and \emph{``context must be rebuilt from scratch every time.''} P2 linked this to output quality: \emph{``generated user stories are too generic because the context is missing.''} Rather than seamless integration, participants describe elaborate manual workarounds. P1 noted: \emph{``there's a huge collection of meeting notes and documentation in Confluence and Jira; however, you have to explicitly instruct Copilot to use it.''} P4 flagged a manual upload process: \emph{``references are currently uploaded manually; in the future, possibly via an interface.''} P8 described the same gap from a different angle: references for tender bids are searched for and adapted using Rovo in Confluence, but must then be transferred manually into TenderZen for the rest of the bid workflow\,--\,a recurring break between two AI tools that each work well in isolation.

Tool proliferation creates its own overhead. P5 observed that \emph{``currently it is necessary to use different tools for different purposes and there are breaks between the tools.''} P7 described a growing toolset where integration efforts are ongoing. While more integrations are constantly added, P7 noted that feedback from the build tool-chain and log file analysis is not yet integrated into the RE tools. P8 illustrated the multi-tool complexity even within a single tender workflow: bid/no-bid decision made in TenderZen, concept development in ChatXiPT and Product Copilot, reference search in Rovo, and requirements transferred to tickets with the help of Product Copilot.

The picture changes when integration does work. P4 reported that TenderZen reduced the evaluation time for a single tender from four hours to 30 minutes. P1 described how bootstrapping a backlog from an Excel spreadsheet was reduced from eight to two hours using Product Copilot. The common factor in these success cases is that AI tools reliably transform unstructured input\,--\,tender documents, spreadsheets, whiteboard photos\,--\,into structured artefacts amenable to further analysis. This transformation is what tools like TenderZen and Product Copilot do well, and it is why integration is the prerequisite for value. Our data captures both ends of this spectrum. P7's adoption of Claude Code connected via MCP to the Jira board and source code repository (see Section~\ref{sec:tooling}) demonstrates how deep integration enables requirements elaboration within the full codebase context. P6, by contrast, works in an Azure DevOps environment where the customer has not enabled MS Copilot's ADO integration and Product Copilot does not support ADO; the result is that P6 has no AI tool operating on the backlog itself.

Integration is not only a technical challenge but also a governance and compliance one. P6 noted: \emph{``Microsoft Copilot has been approved at the customer''}. But P6 has no Pro licence which means that MS Copilot cannot access documents, e.g., in SharePoint. P2 flagged a hard constraint: \emph{``customers do not want to integrate [Product] Copilot due to legal issues.''} P5 voiced similar concerns. Integration thus requires active negotiation around data access and trust. More than 90\% of GenAI-RE approaches remain early-stage prototypes implemented as standalone scripts or web demos~\cite{cheng2026generative}, and systematic reviews on automated RE tools do not treat industrial tool-chain integration as an evaluation dimension~\cite{umar2024advances}. Our data suggest this gap is widening: between late 2025 and spring 2026, practitioners moved from copy-paste workflows to programmatic MCP integrations that the literature has not yet studied.

\paragraph*{AI advances faster than the human and organisational systems around it}

AI tools improve individual PO productivity measurably. P6 noted: \emph{``even this limited use has significantly reduced the effort required to create PBIs; many ideas are clarified in advance that would otherwise only come up during refinement.''} But P5 observed the team-level picture is different: \emph{``the number of refinement loops has not changed due to AI use.''} P1 pointed to team-level opportunities that remain unrealised: they point out the potential for workflow improvement to sort, prioritise, and clean up the backlog, slice epics differently, find duplicates, or archive old stories. AI is primarily being used as a personal productivity tool, not yet as a collaborative team tool.

This points to a design gap rather than a capability gap. The tools we observed treat AI as a personal assistant rather than as a participant in the team's shared refinement process\,--\,in which, as Verwijs and Russo~\cite{verwijs2023theory} show, shared understanding and collective ownership are constituted, not merely communicated. Realising AI's team-level potential would require interaction models that include the whole team: letting developers and other stakeholders see, question, and shape the AI dialogue rather than inheriting only its output. Designing AI into multi-stakeholder team processes rather than individual workflows is precisely the open agenda that Seeber et al.~\cite{seeber2020machines} identify for AI-in-team collaboration; our data show that RE is a domain where this agenda is now pressing.

The technical and communication gap between POs and developers is also shifting in ways that are not yet fully visible. P5 described how quality control has moved to the PO: \emph{``[Product Copilot] can also challenge the stories''} which can mean that the POs together with Copilot can assess technical feasibility without the help of developers. P5 also noted that this disconnect persists in a practical sense when pointing out that Indian developer teams, in other time zones, using different language, could benefit from interacting with the ticket in their language. P7 explicitly raised this as an open question: \emph{``How much does the developer do themselves? Does the PO take on the role for vision and features, rather than user stories?''}

Several participants echo this uncertainty about what it means to be a PO in an AI-assisted world. P1 captured a fundamental question: \emph{``there's no point using [the chatbot] to describe what the story should look like\,--\,if [the] PO can do that, they can write the story directly.''} P2 reframed it at the organisational level: resolving the bottleneck of requirements work means that the development teams need more decision-making autonomy\,--\,and it is not clear that customers are ready for this. As tool integration matures, the ownership dynamic faces additional pressure: if POs can elaborate requirements in the context of the source code without developer involvement, the development team's sense of ownership over the requirements may diminish further\,--\,not because the PO deliberately excludes them, but because the workflow no longer requires their input at the elaboration stage.

Customer and stakeholder readiness is a further constraint that determines what AI assistance can deliver in practice. P6 described an asymmetry: \emph{``UX comes from another company that is hard to reach and has little understanding of how the app works''} This other company does not engage in \emph{``questioning of the use cases\,--\,many features were built for [the app], but the big picture was lost.''} P2 connected this to pace: \emph{``in organisations with less agile thinking and a poor error culture, this is problematic: the Product Owner can produce faster, but cannot necessarily address the technical aspects; the feedback cycle with the customer cannot be accelerated at will.''} The issues around tool approval brought up by P6 show that organisational context limits tool adoption: the tool decision space is constrained by what the customer organisation permits, not just what vendors offer. The human system around the AI tool is the binding constraint. Both the AI-as-teammate literature~\cite{seeber2020machines} and Storey's cognitive-debt framing~\cite{storey2026intentdebt} converge on the same underlying claim: workflows, role definitions, and patterns of communication adapt to AI more slowly than AI use spreads, and the gap is most often closed silently and individually rather than through deliberate redesign. When production accelerates but validation cannot, the risk is not merely wasted effort but misdirected effort: a PO who exceeds the customer's ability to give timely feedback can commit the team to the wrong requirements faster than before. This is a failure mode long recognised where stakeholder communication and shared understanding of requirements break down~\cite{goguen1994requirements,damian2003challenges}.

\paragraph*{Implications for practitioners considering AI-assisted RE}
Our study is descriptive, not prescriptive: we observed eight POs at one service provider company, and we make no claims that what we saw will reproduce elsewhere. Even so, the patterns surfaced in the data suggest a set of questions that practitioners considering AI-assisted RE may find useful to ask of their own situation:

\begin{description}
    \item[On developer--PO interaction:] Do developers actually notice the quality improvements the PO is investing in? What is the team's protocol when AI-generated artefacts arrive in the backlog without the conversational context that shaped them? Does the tooling allow developers and other stakeholders to see and question the AI dialogue, or does the team only inherit its output? Do important communication routes between POs and developers deteriorate when AI is used to shape the technical context of requirements?
    \item[On tool integration:] Which integrations between the team's existing toolchain and any new AI tool are actually in place, which are technically possible but not enabled, and which gaps require manual handoffs that erode the time savings the AI was meant to deliver?
    \item[On organisational lag:] When the PO's productivity goes up, do the team's downstream processes (refinement, planning, customer feedback) absorb that change, or does the PO end up renegotiating workflows and role boundaries silently, one project at a time? Can the customer validate as quickly as the PO can now produce, or does faster output risk committing the team to the wrong requirements?
\end{description}

Our data does not answer these questions for any other organisation; it suggests that they are the right questions to ask early.

\begin{tcolorbox}[width=\textwidth,colback={lightgray},title={Summary of Findings},colbacktitle=darkgray,coltitle=white]
AI-assisted RE tools improve individual PO productivity, but their impact at the team and organisational level remains limited. Three dynamics shape this picture. 
\begin{enumerate}
    \item AI improves story quality, but developers often do not notice the improvement and may resist AI-generated artefacts emotionally; the single-user interaction model means AI assistance substitutes for collaborative dialogue rather than enriching it.
    \item Tool integration is the primary determinant of whether AI value is realised or lost: where integration is in place, effort savings are dramatic; where it is absent\,--\,due to technical gaps, vendor scope, or customer governance decisions\,--\,the PO is left with generic output and manual workarounds.
    \item AI is outpacing the organisational systems around it: benefits accrue to individuals, team processes remain largely unchanged, and customer and stakeholder readiness is more often the limiting factor than tool capability.
\end{enumerate}
\end{tcolorbox}

\section{Threats to Validity}
\label{sec:threats}

Threats to validity follow Runeson and H\"{o}st~\cite{runeson2009guidelines}, adapted for our qualitative interview study.

\paragraph*{Construct validity}
Interviewees may interpret concepts such as ``AI-assisted RE'' or specific tool capabilities differently, potentially introducing inconsistencies in the data. We mitigated this through the semi-structured interview format, which allowed the interviewer to clarify terminology in situ. Member checking ensured that our summaries accurately reflected the interviewees' intended meaning. The codebook itself was derived with AI assistance (Claude Sonnet 4.6) and reviewed manually across several iterations; the resulting code definitions are published as part of the replication package, allowing external readers to verify the constructs to which our findings refer.

\paragraph*{Internal validity}
Since our study is exploratory and descriptive rather than explanatory, we make limited causal claims. Findings could be influenced by confounding factors such as tool maturity, project-specific contexts, or individual PO preferences. The 2024 use-case survey that seeded the initial structural codes could in principle have biased subsequent interview probing; we mitigated this through an emergent thematic-coding pass allowing new themes to surface beyond the survey-derived skeleton, and through purposive sampling that ensured diversity of experience levels (5--18 years), domains, and tool combinations across our eight interviewees. The longitudinal design with two interview rounds (December 2025/January 2026 and April 2026) allowed us to capture how perceptions evolved as tools matured.

\paragraph*{External validity}
Our findings are based on a single B2B software development company with eight POs and a tool landscape covering an in-house chatbot and seven commercial AI tools used for RE. The eight POs work across diverse customer domains (insurance, healthcare, public sector, machinery, e-commerce), which lends some breadth to the observations, but results may not transfer to product companies, larger enterprises, or organisations with different tool ecosystems. We make no generalisability claims beyond our context; the contribution is practical insight into how AI-assisted RE is used and adapted in industry rather than generalisable theory.

\paragraph*{Reliability}
Open coding was performed by the sole author, and interview notes (validated against recordings) rather than full transcriptions served as primary data. As with any single-coder qualitative study, no second, independent human coder applied the codebook, and we therefore cannot report an inter-rater reliability statistic; this is a limitation of our design. Claude Sonnet 4.6 supported the derivation of the coding guidelines and flagged candidate inconsistencies for manual review, and uncertainties were resolved across multiple analysis passes against the interview notes and recordings; this LLM-assisted review improved the internal consistency of code application but does not constitute a validated measure of inter-rater reliability, nor does it substitute for independent human coding. Member checking~\cite{lincoln1985naturalistic} addressed a related but distinct concern\,--\,whether our interpretations matched participants' intended meaning\,--\,rather than coding reliability. We partially offset the absence of IRR by making the interview guide and the full coding schema available as supplementary material~\cite{supplementary-material}, allowing external verification of how codes were defined and applied.

\section{Conclusion}
\label{sec:conclusion}

AI-assisted requirements engineering in practice reveals a picture with distinct sociotechnical elements. Across eight product owners and 15 use cases at XITASO, a medium-sized software engineering firm, the influence of AI lies not in task automation alone but in how it reshapes the conversations, roles, and team dynamics through which requirements acquire meaning. The AI-augmented PO is outpacing the team and the customer: benefits accrue to individuals while team processes, customer readiness, and organisational workflows absorb change more slowly than the tools themselves are being adopted.

This has a theoretical edge. Requirements engineering has long been understood as a distributed, negotiated activity in which validation and shared understanding are constituted through social interaction, not merely transmitted via artefacts~\cite{goguen1994requirements,damian2003challenges}. The single-user interaction model of every tool we observed relocates this work into a private PO--AI conversation, producing artefacts the team inherits without the reasoning that shaped them. Agile role boundaries and the collaborative construction of requirements do not remain stable once AI enters the workflow: the boundary between requirements and implementation blurs, and the question of who owns a requirement becomes open. In doing so, our findings extend Storey's debt framework~\cite{storey2026intentdebt} into the RE domain: intent debt at the PO--AI layer, where unreviewed artefacts carry no captured rationale, and cognitive debt at the PO--team layer, where the team's loss of transactive memory around AI-mediated conversations erodes shared understanding. Together they provide the RE-domain evidence that the AI-as-teammate literature has called for~\cite{seeber2020machines,webber2024paradox}.

The risk is not that AI produces poor requirements, but that it redistributes where requirements work happens without anyone deciding that it should. Unreviewed AI artefacts reaching developers, conversational context that never leaves the PO's screen, and developers' diminishing sense of ownership together accumulate as unverified requirements and eroded shared understanding. Because the customer feedback cycle cannot be accelerated at will, a PO who produces faster than the team can absorb and faster than the customer can validate risks misalignment rather than inefficiency: delivering the wrong thing at higher speed.

Addressing this requires more than studying AI-assisted RE over longer horizons. It calls for interaction models and tools designed for multi-party rather than single-user use, and for evaluations that treat tool-chain integration and organisational absorption as first-class concerns. For practitioners, it means deliberately redesigning refinement rituals and role boundaries around AI, deciding consciously how the team re-enters the loop, rather than letting those adjustments happen piecemeal and out of sight.

\section*{Data Availability} 
Due to participant confidentiality agreements and the proprietary information about customers and internal practices included in the interviews, interview transcripts cannot be shared publicly. However, the following materials are available in the supplementary material~\cite{supplementary-material}: (1) the interview protocol with all questions, (2) the complete coding schema with code definitions and examples, (3) the complete electronic whiteboard with all anonymized data points and attached codes, and (4) an overview of all use cases extracted from the data.

\bibliography{ai-assisted-re}

\end{document}